\documentclass[referee]{aa}

\usepackage{graphicx}
\begin{document}
\bibliographystyle{plainnat.bst}

\voffset+1.5cm
\newcommand{\gsim}{\hbox{\rlap{$^>$}$_\sim$}}
\newcommand{\lsim}{\hbox{\rlap{$^<$}$_\sim$}}

\title{Conical Fireballs, Cannon Balls And Jet-Breaks\\
 In The Afterglows Of Gamma Ray Bursts}

\author{Shlomo Dado \and Arnon Dar}
\institute{Physics Department, Technion, Haifa 32000, Israel}

\date{}

\abstract{The 'jet-break' in the X-ray afterglow of gamma ray bursts
(GRBs) appears to be correlated to other properties of the X-ray
afterglow and the prompt gamma ray emission, but the
correlations are at odds with those predicted by the conical fireball (FB)
model of GRBs. They are in good agreement, however,  with those
predicted by the cannonball (CB) model of GRBs.}


\maketitle

\titlerunning{Jet Breaks In The  Afterglow Of GRBs}
\authorrunning{Shlomo Dado \& Arnon Dar}

{\centering 
{\bf Conical Fireballs, Cannon Balls And Jet-Breaks\\
In The Afterglows Of Gamma Ray Bursts}\\

\centering
{Shlomo Dado and Arnon Dar}\\

\centering
{Physics Department, Technion, Haifa 32000, Israel}\\

\centering 
{{\bf abstract.} The 'jet-break' in the X-ray afterglow of gamma ray 
bursts (GRBs)\\
appears to be correlated to other properties of the X-ray
afterglow and the prompt\\
gamma ray emission, but the correlations are at odds with those predicted 
by the\\ 
conical fireball (FB)
model of GRBs. They are in good agreement, however, with\\
the correlations that are 
predicted by the cannonball (CB) model of GRBs.}\\}

\section{Introduction}

Before the launch of the Compton Gamma Ray Observatory (CGRO) in 1991 it 
was widely believed that gamma-ray bursts (GRBs) originate in the Galaxy 
or in its halo. Much larger distances together with the observed fast 
rise-time of GRB pulses implied an 'energy crisis' - implausible energy 
release in gamma rays from a very small volume in a short time, if the 
emission was isotropic, as was generally assumed. However, the isotropic 
distribution of GRBs over the sky and their intensity distribution that 
were measured with the Burst and Transient Source Experiment (BATSE) 
aboard the Compton Gamma Ray Observatory (CGRO) shortly after its launch 
provided a clear evidence that the observed GRBs are at very large 
cosmological distances (Meegan et al.~1992; Mao and Paczynski~1992). That 
led Shaviv and Dar (1995) to propose that GRBs are produced by inverse 
Compton scattering of light by highly relativistic jets whose radiation is 
narrowly beamed along their direction of motion and that such jets 
presumably are ejected following violent stellar processes such as core 
collapse supernovae, merger of compact stars, mass accretion on compact 
stars and phase transition in compact stars, rather than in spherical 
fireballs (Paczynski~1986, Goodman~1986) produced by neutron star merger 
in close binaries (Goodman et al.~1987). These jets were assumed to be a 
succession of highly relativistic plasmoids of ordinary matter like those 
observed in high resolution observations of highly relativistic jets 
ejected from galaxies with active galactic nucleus (e.g., M87 in Virgo), 
from radio galaxies (e.g., Centaurus A, the nearest radio galaxy), from 
quasars (e.g., 3C 273 in Virgo) and from microquasars (e.g., GRS 1915+105, 
SS 433, Cygnus X-1 and cygnus X-3 in our Galaxy). A key prediction of such 
a model was a very large linear polarization ($P\sim 100\%$) of gamma rays 
observed from the most probable viewing angle of the GRBs (Shaviv and Dar 
1995).

The hypothesis that GRBs are produced by highly relativistic jets was not 
widely accepted even when the discovery of the X-ray afterglows of GRBs by 
the Beppo-SAX satellite (Costa et al.~1997)  allowed their arcminute 
localization that led to the discovery of their longer wave-length 
afterglows (van Paradijs et al.~1997, Frail et al.~1997), host galaxies 
(Sahu et al.~1997) and their large redshifts (Metzger et al.~1997).  In 
fact, the discovery 
of GRB afterglows, which appeared to decay like a single power-law in 
time, as predicted by Paczynski and Rhoads (1993), Katz (1994) and Mezaros 
and Rees (1997) from the isotropic fireball (FB) model (Paczynski~1986, 
Goodman~1986), led to an immediate wide acceptance of the relativistic 
isotropic fireball model as the correct description of GRBs and their 
afterglows (see, e.g., Wijers et al.~1997, Piran~1999), ignoring the 
'energy crisis' of the isotropic fireball model of GRBs.

When data on redshifts and afterglows of GRBs began to accumulate, it 
became 
clear that GRBs could not be explained by the isotropic fireball model 
(e.g., Dar~1998).  Not only their large redshifts implied implausible 
energy release in gamma rays if the emission was isotropic, 
their observed afterglows seemed to behave like a smoothly broken 
power-law in time (Beuermann et al.~1999; Fruchter et al.~1999; Harrison 
et al.~1999; Kulkarni et al.~1999) rather than a single power-law. Only 
then, the isotropic fireball was replaced (e.g., Sari et al.~1999; Piran 
2000) by an assumed  conical jet of thin shells where synchrotron 
radiation from collisions between overtaking shells (or internal shocks) 
produce the observed GRB pulses, and the following collision of the 
merged shells with the interstellar medium (ISM) produces the synchrotron 
afterglow. This conical jet model which was given the name 'collimatedā 
fireball model', replaced the original fireball model, but retained 
its name  'the fireball model'.

The afterglow of a conical shell of opening angle $\theta_j$, whose 
propagation is decelerated by 
sweeping up the interstellar medium (ISM) in front of it, was shown by 
Sari et al.~1999 to have an {\it achromatic break} when its bulk motion 
Lorentz factor $\gamma(t)$ has dropped below the value 
$\gamma(t)=1/\theta_j$, which was argued to be roughly at the transition 
of the jet from a cone-like shape to a trumpet-like shape due to the 
lateral expansion of the conical jet. Moreover, the conical fireball model 
has been used to predict the pre-break and post-break temporal and 
spectral indices of the spectral energy density $F_\nu(t)\propto 
t^{-\alpha}\, \nu^{-\beta}$ of the afterglow and the {\it closure 
relations} that they satisfy.

Because of the complexity of the dynamics of spreading jets and the 
dependence of their afterglow on many adjustable parameters, the observed 
afterglows of GRBs rarely have been modeled with theoretical light-curves 
calculated from the conical fireball model. In most cases they were fitted 
with heuristic sharply or smoothly broken power-law functions connecting 
the pre-break and post-break behaviours predicted by Sari et al.~(1999). 
Such heuristic functions were used primarily for convenient 
parametrization of the data. They allowed, however, to extract a 
break-time $t_b$ from the observed light-curve and to test whether the 
pre-break and post break slopes satisfy the closure relations of the 
conical fireball model.

The jet breaks, however, were found to be chromatic (e.g., Covino et al. 
2006, Panaitescu et al. 2006). The X-ray afterglows of GRBs with large 
equivalent isotropic energy ($E_{iso}\gg 10^{53}$ erg) that were observed 
with the Swift X-ray telescope (XRT) (e.g., GRBs 061007, 130427A) showed a 
single power-law behaviour with no visible jet break, and almost all the 
X-ray afterglows of less energetic GRBs that appeared to have a 'jet 
break' did not satisfy the closure relations of the conical fireball 
model, either before the break or after it (e.g., Liang et al.~2008; 
Racusin et al.~2009). In particular, a large fraction of the X-ray 
afterglows of GRBs measured with the Swift X-ray telescope (Swift/XRT) 
showed a canonical behaviour (Nousex et al.~2006) where the afterglow has 
a shallow decay phase (plateau) before the break with $\alpha(t<t_b)\ll 1$ 
far from the predicted $\alpha_X(t<t_b)=(3\,\beta_X-1)/2$. Despite these 
failures and many other failures of the fireball model, the model has not 
been given up. Instead, the missing breaks were attributed to various 
reasons such as quality of the data (Curran et al.~2008), break-time 
beyond the end of the Swift/XRT follow-up observations (Kocevski and 
Butler~2008) and far off-axis observations (Van Eerten et al.~2011a). The 
failure of the pre-break closure relation, to describe the shallow 
decay/plateau phase of canonical X-ray afterglows was attributed to an 
assumed continuous energy injection. The chromaticity of the jet break and 
the failure of the closure relation for the post-break behaviour of the 
X-ray afterglow were largely ignored.

In order to test whether part of the above difficulties arise from 
approximations used in the analytical calculations, and in order to 
generalize the predictions to off-axis observers, various authors have 
tried to derive the light-curves of conical fireballs from numerical 
hydrodynamical calculations. In particular, recently van Eerten and 
MacFadyen (2012) reported two dimensional (2D) numerical hydrodynamic 
calculations of the light-curves of the afterglow from conical fireballs 
observed from an arbitrary angle. These numerical calculations showed that 
the difference in the temporal indices across the jet break is larger than 
that predicted by Sari et al.~(1999) and, contrary to expectation, it 
increases the discrepancy between theory and observations rather than 
removing it. The pre-break behaviour remains an unsolved difficulty, which 
was speculated to be due to an assumed continuous energy injection 
into the conical fireball. It was also speculated that the discrepancy 
between the post-break temporal slopes obtained from the numerical 
simulations and those observed with the Swift/XRT may be removed or 
reduced by assuming that the afterglow is produced by a blast wave that 
decelerates in a wind environment rather than in a constant density ISM.

All the above difficulties of the conical fireball (FB) model, however, 
were  not shared by the cannonball (CB) model of GRBs:  The 
canonical behaviour of X-ray afterglows where a plateau/shallow decay phase 
is smoothly broken to a steep power-law decline was predicted long before it 
was observed with Swift/XRT (see, e.g., Figs. 6, 26-30 in Dado et 
al.~2002; see also Dado et al.~2009a,b for a detailed comparison between 
the light-curves of the X-ray afterglows of GRBs measured with Swift/XRT 
and those predicted by the CB model). The post-break closure relations 
predicted by the CB model were also shown to be well satisfied by the 
Swift/XRT light curves (Dado and Dar~2012a).

The failures of the standard conical fireball model, however, did not 
appear to shake the wide belief in this model or in its interpretation of the 
afterglow breaks. Hence, in this paper we present additional {\it 
parameter-free} tests of the origin of the observed break in the 
light-curve of canonical  X-ray afterglows of GRBs. Namely, we compare the 
observed correlations between the jet breaks in the X-ray afterglows of 
GRBs measured with the Swift/XRT between December 2004 and December 2012 
and the prompt gamma-ray emission properties of these GRBs, and those 
predicted by the conical fireball and cannonball models. We limit our 
tests to the X-ray afterglow, in order to avoid dependence on adjustable 
parameters. This extends our preliminary study of missing breaks in the 
X-ray afterglows of GRBs (Dado et el~2007) that was based on limited 
statistics. For completeness, the derivation of the break properties from 
the conical fireball model and from the cannonball model are presented in 
Appendixes A1 and A2, respectively.

\section{Jet Break Correlations}

\subsection{Conical jet break}

In standard conical jet models of GRBs, $1/\gamma(0)\ll \theta_j$. The 
break in the 
afterglow occurs when the beaming angle $1/\gamma(t)$ of 
the emitted 
radiation from the decelerating jet in the interstellar medium (ISM) 
becomes larger than the opening angle $\theta_j$ of the conical jet, 
i.e., when $1/\gamma(t_b)\approx \theta_j$. For a conical jet at redshift $z$ 
with kinetic energy $E_k$ propagating in an ISM with a constant baryon 
density 
$n_b$, the break is observed by a distant observer on/near axis at a time 
(see Appendix A)
\begin{equation}
t_b\approx {(1+z)\over 8\,c}\,\left[{3\,E_k\over 2\,\pi\, n_b\, 
m_p\,c^2}\right]^{1/3}\,\theta_j^2.
\label{tbfb}
\end{equation}
Eq.~(1) is the relation derived by Sari 
et al.~(1999)  for a conical shell which begins rapid  lateral spreading
on top of its radial motion when $\gamma(t)\approx 1/\theta_j$. 

If the jets that produce GRBs had approximately a standard ISM environment 
and a standard $E_k$ (Frail et al.~2001), then Eq.(1) 
would have yielded the correlation
\begin{equation}
t'_b \propto [E_{iso}]^{-1}\,,
\label{tbc1}
\end{equation}
where $E_{iso}$ is the total gamma-ray energy emission under the 
assumption of isotropic emission.

Because $1/\gamma(0)\ll \theta_j$, the highly relativistic conical ejecta 
and its beamed gamma-ray emission share the same cone.  Since the observed 
spectrum of GRBs is given roughly by a cutoff power-law (CPL) 
$E\,dn_\gamma /dE\propto e^{-E/E_p}$, the assumption of the conical 
fireball model that a constant fraction of the jet kinetic energy is 
converted to gamma-ray energy implies that
\begin{equation}
E'_p\propto {E_k \over \pi\,\theta_j^2} \propto E_{iso}\,,
\label{EpEiso1}
\end{equation}
where $E'_p$ is the peak energy of the time-integrated observed spectral 
energy flux. Consequently, the fireball model assumptions yield also the 
binary correlation $t'_b \propto [E'_p]^{-1}.$

Note also that for a ballistic (non spreading) conical jet viewed 
on/near-axis, the power-law decline of the X-ray afterglow of an 
isotropic fireball seen by a distant observer is multiplied by a factor 
\begin{equation} 
K_b= \left[ 1- {1\over (1+\gamma^2\,\theta_j^2)^{\beta_X+1}}\right]\,, 
\label{Kb} 
\end{equation} 
which follows from Eq.~(A.6)  of Appendix A1 for $\theta_j^2\ll 1$.  For 
$t\ll t_b$,   $\gamma^2 \theta_j^2\gg 1 $ and $K_b\approx 1$, while 
for  $t\gg t_b$,   $\gamma^2 \theta_j^2\ll 1 $ and 
$K_b\approx (\beta_X+1)\,\gamma^2 \theta_j^2\propto t^{-3/4}$.
Hence, the temporal index $\alpha$ of the afterglow of a conical jet 
increases by $\Delta \alpha=0.75$ across the break, independent of the 
spectral index $\beta_X$ and the pre-break temporal index of the afterglow. 

In the case of a wind-structured environment with a density profile $n(r)= 
n_0\,R_0^2/R^2$, the power-law indices of the $t'_b-E_{iso}$ and 
$t'_b-E'_p$ correlations are identical to those for an ISM circumburst 
environment (see Appendix A1).

Moreover, although the $t'_b-E_{iso}$ and $t'_b-E'_p$ correlations were 
derived for the case where no continuous energy injection during the 
plateau phase of the X-ray afterglow takes place, they are valid also, to 
a good approximation, when continuous energy injection is invoked. This is 
because the injected energy during the afterglow phase phase must be much 
smaller than the initial kinetic energy which powers the prompt gamma-ray 
emission whose energy is is much larger than that of the afterglow. Since 
the afterglow is only partially powered by the assumed continuous energy 
injection, the continuous energy injection must be rather small compared 
to $E_k$, the total kinetic energy of the jet. Hence, the assumption, 
$E_{iso}\propto E_k$ holds to a good approximation and Eq.~(A.1) implies 
that the correlations derived from Eq.~(A.1) are valid also when such a 
continuous energy injection after the prompt emission phase is present.

\subsection{Cannonball deceleration break}

In the CB model of GRBs a succession  of initially   
expanding plasmoids (CBs) of ordinary matter merge into a slowly expanding 
($(kT/m_p\,\gamma^2)^{1/2} \ll c$) leading 
CB with a large bulk motion Lorentz factor, $\gamma(0)\sim 10^3$, that 
decelerates in 
collision with the circumburst medium/ISM. The emitted
synchrotron radiation is relativistically beamed along its direction of 
motion, redshifted by the cosmic expansion and its arrival time 
in the frame of a distant observer at a viewing  angle $\theta$
relative to the direction of motion of the CB is aberrated (e.g.,
Dar and De R\'ujula 2004 and references therein). The rate of change 
in the bulk motion Lorentz and Doppler factors of the CB 
due to the deceleration of the CB in the circumburst medium
is small until the swept-in mass by the CB becomes comparable to its 
initial mass. This happens in the observer frame at a time 
(see, e.g., Dado et al. 2009a and references therein)
\begin{equation}
t_b \approx {(1+z)\, N_b\over 8\,c\, n_b\,\pi\, R^2}\, 
{1\over \gamma_0\,\delta_0^2}\,,
\label{tbreakcb}
\end{equation}
where  $N_ b$ is the baryon number of the CB and $R$ is its radius. 
The rapid decrease of $\gamma(t)$ and $\delta(t)$ beyond $t_b$
produces a smooth transition (break) around $t_b$ 
of the observed spectral energy density
$F_\nu(t) \propto [\gamma(t)]^{3\,\beta-1}\, [\delta(t)]^{\beta+3}\, 
\nu^{-\beta}$ of the emitted afterglow from 
a plateau phase to an asymptotic power-law decline
(e.g., Dado et al.~2002,2009a and references therein). 
 
In the CB model, $E'_p\propto \gamma_0\, \delta_0$ and $E_{iso} 
\propto 
\gamma_0\, \delta_0^3$, respectively. Consequently, Eq.~(5) yields the 
triple correlation, 
\begin{equation} 
t'_b \propto [E'_p\, E_{iso}]^{-1/2}\,. 
\label{tbreakepeiso} 
\end{equation} 
Moreover, substituting the CB model approximate binary 
power-law correlation $E'_p \propto [E_{iso}]^{1/2}$ into the triple 
$t'_b-E'_p-E_{iso}$ correlation yields the $t'_b-E_{iso}$ and $t'_b-E'_p$ 
binary correlations, 
\begin{equation} 
t'_b \propto [E_{iso}]^{-3/4}\propto [E'_p]^{-3/2}\,,
\label{tbreakeiso} 
\end{equation} 
where a prime indicates the value in the GRB rest frame. 
Naturally, these approximate correlations are expected to have a larger 
spread than the original triple $t'_b-E'_p-E_{iso}$ correlation.

In the case of a wind-like density, which extends typically beyond the end 
of the glory region at $R_g\sim 10^{16}$ cm (Dado et al.~2009) and up to 
$R_w \sim 5\times 10^{17}$ cm where the wind density decreases below 
$\sim m_p/{rm cm^3}$ (for the wind parameters listed in Appendix A1), 
the 
predicted spectral energy flux of the afterglow has the behaviour 
$F_\nu(t)\propto t^{-(\beta+1)}\,\nu^{-\beta}$ where $\beta(t)$ is the 
spectral index in the observed band (see, e.g., Dado et al.~2009). The 
observed crossing time of such a wind region is roughly 
$(1+z)\,R_w/\gamma_0\, \delta_0$, typically $< 50\,(1+z)$ s  because 
$\gamma$ and $\delta$ change little during the wind crossing. Beyond the 
wind region, $F_\nu(t)$ in the X-ray band has the standard 
canonical behaviour of X-ray afterglows in the CB model in an ISM 
environment, i.e., with an afterglow  break/bend  at the end of a plateau 
phase that satisfies Eq.~(7).

\section{Comparison with observations}

Fig.~1 presents the best fit power-law to the observed 
$E'_p-E_{iso}$ correlation using  an unbiassed  
sample of 110  GRBs  with known redshift measured before January 1, 2013.
The values of $E'_p$ and $E_{iso}$ were
compiled  from  communications of the
Konus-Wind and Fermi GBM collaborations
to the  GCN Circulars Archive (Barthelmy 1997), and from
publications by Amati et al.~(2007,2008), Yonetoku et al.~(2010),   
Gruber et al.~(2011),  Nava et al.~(2012) and D'Avanzo et al.~(2012).
Using essentially the method advocated by  D'Agostini~(2005), 
we obtained the best fit power-law correlations
$E'_p \propto  [E_{iso}]^{0.54}$ in 
good agreement with $E'_p \propto  [E_{iso}]^{1/2}$
predicted by the CB model but in  disagreement with 
$E'_p \propto E_{iso}$ expected in the  conical fireball 
model.
\begin{figure}
\centering
\resizebox{\hsize}{!}{\includegraphics{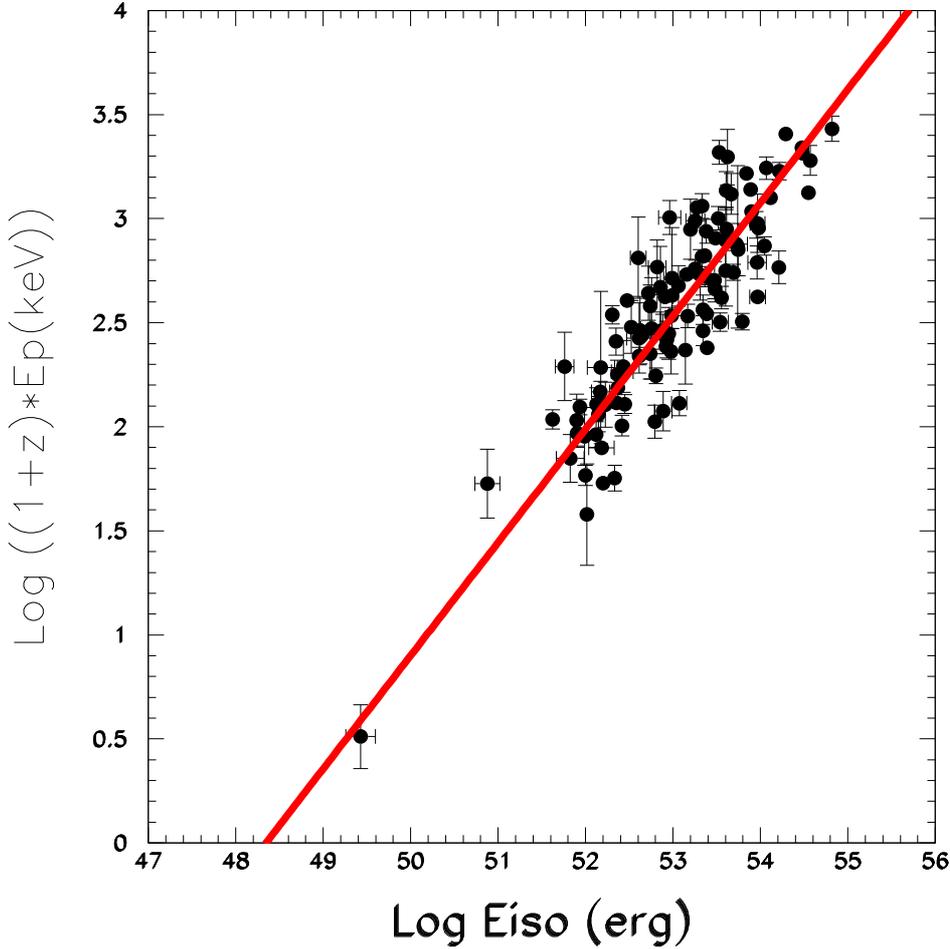}}
\caption{The observed correlation between $E'_p$ and $E_{iso}$ 
for 121 long GRBs with known redshift. The best fit power-law correlation 
(straight line) has a power-law index 0.54.}
\label{FIGL1}
\end{figure}

Fig.~2 compares the triple correlation $t'_b-E'_p-E_{iso}$ predicted by 
the CB model (Eq.~6) and the observed correlation in 70  Swift GRBs (Evans 
et al.~2009) from the above GRB sample, which have a good Swift/XRT 
temporal sampling of their X-ray afterglow during the first day (or more) 
following the prompt emission phase and have no superimposed flares. In 
this sample the X-ray afterglow of 55 GRBs clearly show a break and no 
afterglow-break was observed in 15 GRBs.  The upper bound on a 
possible early time break for the 15 GRBs with no visible break are 
indicated by down pointing arrows. Also shown is the late-time break 
of the X-ray afterglow of GRB 980425, which was measured with Chandra 
(Kouveliotou et al.~2004). In order not to 
bias the values of $t_b$ by the CB model fits, the break times  
were taken to be the times of the first break with 
$\alpha(t<t_b)<\alpha(t>t_b)$ obtained from a broken power-law fit to the 
GRB X-ray afterglow measured with the Swift/XRT and reported in the 
Leicester XRT GRB catalog (Evans et al.~2009) or from their smoothly 
broken power-law fits of Margutti et al.~(2013). The Spearman rank 
(correlation coefficient) of the triple correlation $t'_b-(E'_p\,E_{iso})$ 
for the subsample of $55$ GRBs with a visible break is $r= -0.74$ 
corresponding to a 
chance probability less than $1.4\times 10^{-10}$. The best fit 
triple 
correlation $t'_b \propto [E'_p\, E_{iso}]^q $ that was obtained for the 
subsample of 55 GRBs using essentially the maximum likelihood method 
advocated by D'Agostini~(2005), yields $q=-0.58\pm 0.04$.

\begin{figure}
\centering
\resizebox{\hsize}{!}{\includegraphics{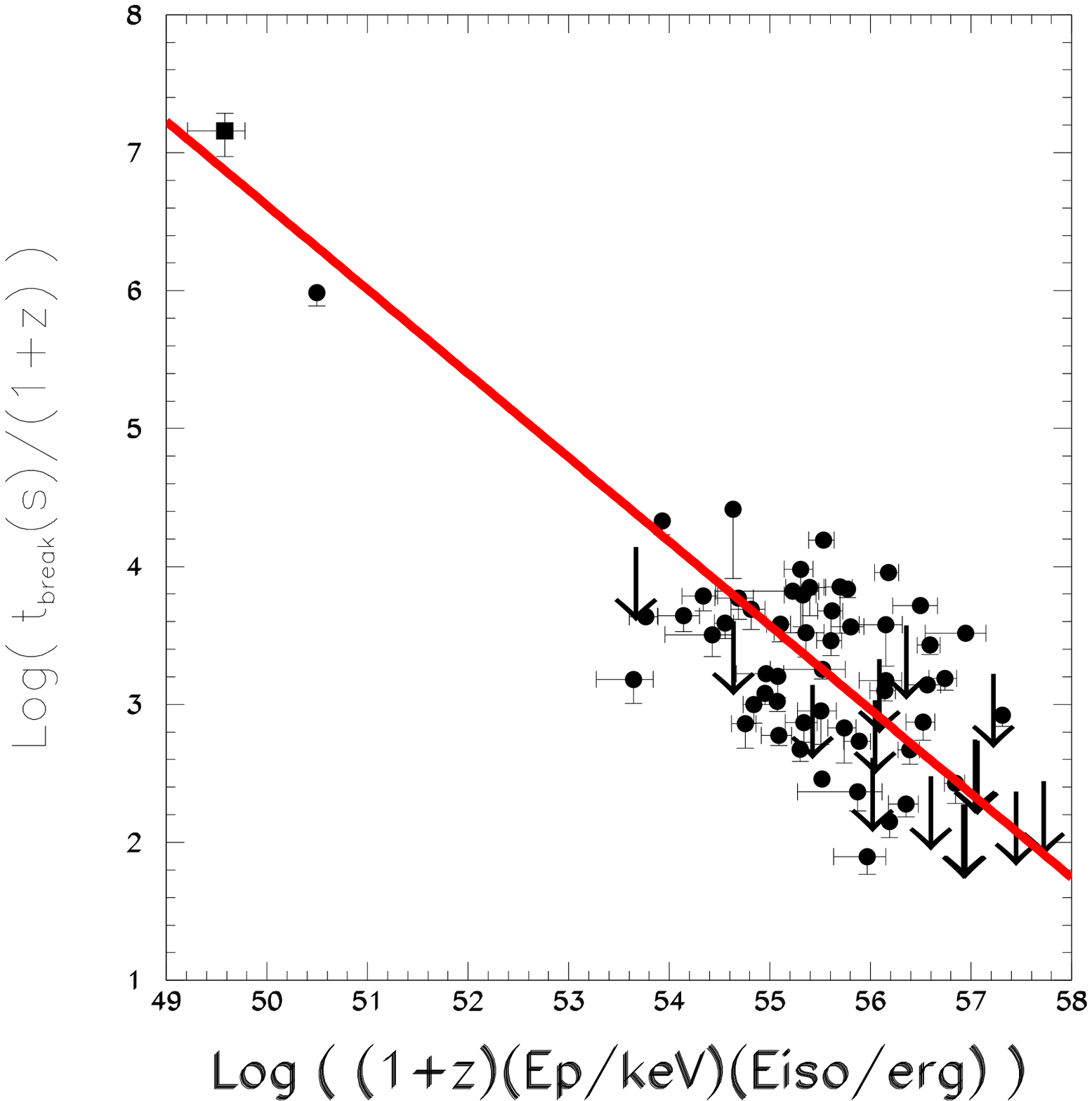}}  
\caption{The observed triple correlations
$t'_b-E'_p-E_{iso}$
in 70 Swift GRBs with measured  redshift,
$t'_b$,  $E'_p$, and $E_{iso}$ and its best fit power-law
(straight line with a power-law index -0.58). Arrows
indicate observational upper bounds on early-time deceleration breaks
before the beginning of the Swift/XRT observations or hidden under the
prompt emission tail. The square represents the break in the 
late-time X-ray afterglow of GRB 980425 which was 
observed with Chandra.}
\label{FIGL2}
\end{figure}

The approximate binary correlations $t'_b-E_{iso}$ and $t'_b-E'_p$ that 
were obtained by substitution of the CB model predicted correlation $E'_p 
\propto [E_{iso}]^{1/2}$ in the triple correlation $t'_b-E'_p-E_{iso}$ 
(Eq.~6) are compared with the observational data in Figs.~3 and 4, 
respectively. GRB 980425 was excluded from the GRB sample because in the 
CB model the $E'_p \propto [E_{iso}]^{1/2}$ is satisfied only by ordinary 
GRBs where $\theta\approx 1/\gamma$, while far-off axis GRBs such as 980425 
with $\theta\gg 1/\gamma$ satisfy $E'_p\propto [E_{iso}]^{1/3}$, i.e., 
they 
are outliers with respect to the assumed $E'_p \propto [E_{iso}]^{1/2}$ 
correlation. The Spearman ranks of the observed $t'_b-E_{iso}$ and 
$t'_b-E'_p$ correlations are -0.49 and -0.63 with chance 
probabilities less than $4.5\times 10^{-4}$ and $1.0\times 10^{ -6}$, 
respectively, and as expected (in the CB model), they are larger than that 
of the $t'_b-(E'_p\, E_{iso})$ correlations. The best fit power-law 
indices of the $t'_b-E_{iso}$ and $t'_b-E'_p$ correlations are $-0.70\pm 
0.06$ and $-1.64 \pm 0.04$ , respectively, consistent with their predicted 
values by the CB model, $-0.75$ and $-1.50$, respectively.

\begin{figure}
\centering
\resizebox{\hsize}{!}{\includegraphics{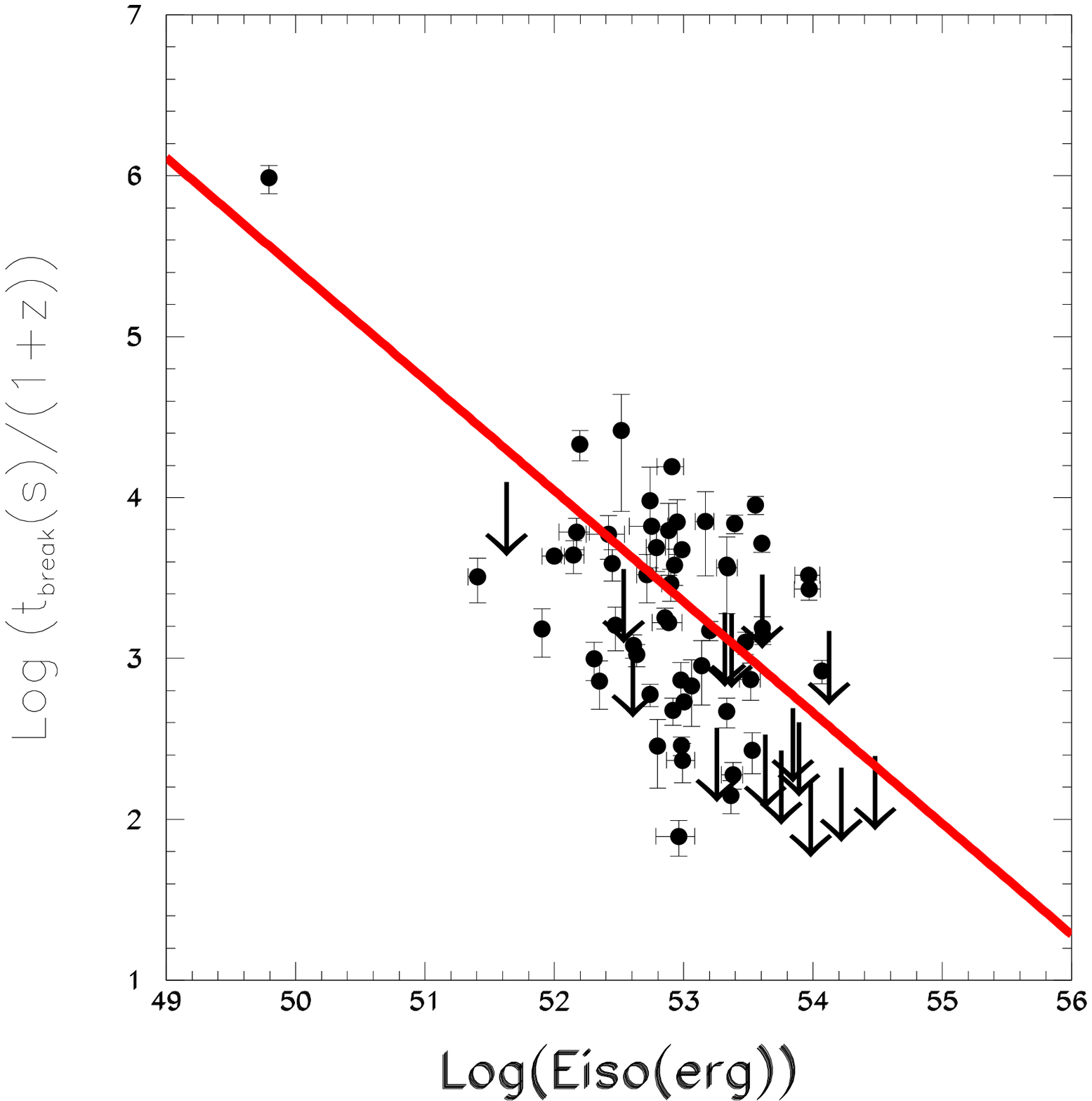}} 
\caption{Comparison between the binary correlation
$t'_b-E_{iso}$
predicted by the CB model (Eq.~7) and that observed in
70 Swift GRBs with known redshift,
$t'_b$ and $E_{iso}$. Arrows indicate
observational upper bounds on early time deceleration breaks
before the beginning of the Swift/XRT observations or hidden under the
prompt emission tail.}
\label{FIGL3}
\end{figure}

\begin{figure}
\centering
\resizebox{\hsize}{!}{\includegraphics{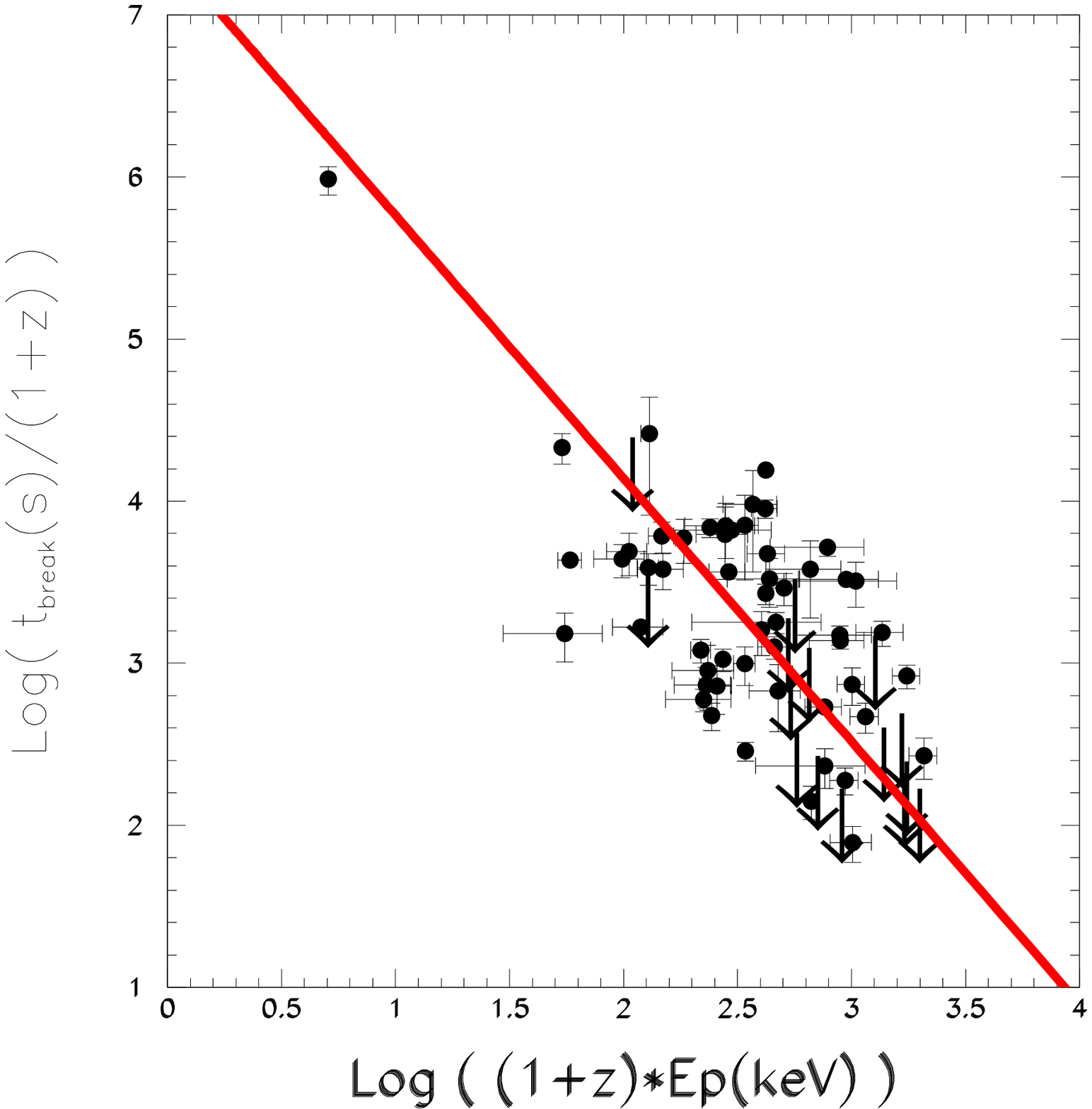}}
\caption{Comparison between the binary correlation
$t'_b-E'_p$ predicted by the CB model (Eq.~7) and that observed in
70 Swift GRBs with known redshift,
$t'_b$ and $E'_p$. Arrows indicate
observational upper bounds on early time deceleration breaks which 
may have taken place
before the beginning of the Swift/XRT observations or are hidden under the
prompt emission tail.}
\label{FIGL4}
\end{figure}

A comparison between the best fit indices of the break-time power-law 
correlations and those expected in the CB and FB models is summarized in 
Table 1 for the sample of 55 Swift GRBs with a visible afterglow break. As 
can be seen from Table 1, the values of the power-law correlation indices 
predicted by the CB model are consistent with those obtained from the best 
fits.  The best fit indices, $0.54\pm 0.01$ , $-0.69\pm 0.06$ and 
$-1.62\pm 0.04$ of the observed $E'_p-E_{iso}$, $t'_b-E'_p$ and 
$t'_b-E_{iso}$ power-law correlations, however, are at odds with the 
values 1, -1, and -1, respectively, expected in the conical fireball 
model. While the $\chi^2/dof$ of the predicted correlations by the CB 
model differ from those of the best fits by less than 1/dof, the 
$\chi^2/dof$ of the predicted correlations by the FB model differ by much 
larger  values,  as summarized in Table 2.

The correlations satisfied by $t'_b$ imply that GRBs with very large 
$E_{iso}$ and/or $E'_p$ have a break at small $t'_b$, which is hidden 
under the tail of the prompt emission or precedes the start of the 
Swift/XRT follow up observations (Dado et al. 2007). Indeed, the X-ray 
afterglow of all the 15 GRBs in the sample, which have a very large 
$E_{iso}\times E'_p$, have a single power-law decline consistent with the 
post-break power-law decline predicted by the CB model (see, e.g., Dado et 
al.~2007, 2009a). For such GRBs, the observations provide only upper 
bounds for the break-times of their X-ray afterglows. The correlations 
satisfied by $t'_b$ imply that GRBs with very large $E_{iso}\times E'_p$ 
have a break at small $t'_b$, which is hidden under the tail of the prompt 
emission or precedes the start of the Swift/XRT follow up observations 
(Dado et al. 2007). Indeed, the X-ray afterglow of all the 15 GRBs in the 
GRB sample, which have a very large $E_{iso}$, and/or $E'_p$, have a 
single power-law decline consistent with the post-break power-law decline 
predicted by the CB model (see, e.g., Dado et al.~2007, 2009a).  For such 
GRBs, the observations provide only upper bounds for the break-times of 
their X-ray afterglows.

However, for most of the 15 GRBs with only an upper bound on their 
afterglow  break-time,  
which are indicated by down pointing errors in Figs.~2-4, an early 
break-time value was  extracted from a CB model fit to the entire 
X-ray light curve, which includes the fast decline phase of the prompt 
emission and the afterglow component (see, e.g., Dado et al. 2009). 
Replacement of the upper bounds by the CB model fitted 
break-times (which could have biased the break-times values) and their 
inclusion in the best fits had very little effect on the values of the 
best fit power-law indices and their errors.

\section{Conclusions and discussion}

Correlations and closure relations between GRB properties that are 
predicted by GRB models allow parameter-free tests of such models. In 
particular, comparison between the observed and predicted correlations 
between the 'jet break' in the X-ray afterglow of GRBs and 
the prompt 
gamma-ray emission, like the $E'_p-E_{iso}$ correlation,
allow another critical test of the conical fireball 
model and the cannonball model of GRBs. Although the 'jet break' in the 
afterglow of GRBs has been the flagship of the conical fireball model, the 
observed correlations between the 'jet break' in the X-ray afterglow of 
GRBs measured with the Swift/XRT and their prompt gamma-ray emission are 
inconsistent with those expected in the conical fireball model.  This 
failure, perhaps is not a surprise since the observed 'jet-breaks' were 
found before 
to be chromatic, the predicted pre-break and post-break temporal 
behaviours and closure relations were found to be badly violated, and the 
observed change in slope across the breaks is not that predicted. The 
replacement of the approximate analytical estimates in the conical 
fireball model (Sari et al. 1999) by more exact hydrodynamical 
calculations (e.g. van Eerten and MacFadyen 2012) does not change the 
situation. They neither reproduce the observed correlations, nor do they 
remove the discrepancies between the predicted and observed pre-break and 
post break behaviours of the afterglows. These failures provide additional 
evidence that GRBs and their afterglows are not produced by  conical 
jets, the so called 'collimatedā fireballs'.

In contrast, the correlations between the deceleration break in the 
afterglow of GRBs and their prompt $\gamma$-ray emission 
predicted by the cannonball model are in good agreement with those 
observed, as shown in Figs 1-4. The correlations between the break and 
other afterglow properties predicted by the cannonball model (Dado and 
Dar~2012b), as well as the pre-break and post-break behaviours of the 
light curves of the X-ray afterglow, were 
shown to accord well with the
observations (e.g., Dado et al. 2009, Dado and Dar 
2012a). Moreover, in the CB model, the $t'_b-(1+z)$ 'anti-correlation' 
noted by Stratta et al.~(2009), is a simple consequence of 
beaming and the detection threshold, which enrich the low $z$ GRB sample 
with far-off-axis soft GRBs and X-Ray Flashes (e.g., Dado et al.~2004)  
relative to the high $z$ events that must be much harder and energetic in 
order to be detected. These selection effects that produce
the effective $<E'_p(z)>-z $ and $<E_{iso}(z)>-z$ 'correlations'
result in an  effective  $t'_b(z)-z$ 'anti-correlation'
(Dado and Dar, in preparation).

\begin{table}
\caption{Summary of the observed power-law correlations 
between $E'_p$, $E_{iso}$ and $t'_b$ and their power-law indices    
expected in  
the Cannonball (CB) and collimated fireball (FB) models. $\rho$ is the  
Spearman rank 
(corrrelation coefficient), P$(\rho)$  is  the chance probability of 
a correlation coefficient $\geq \rho$,
and $p$ is the power-law index of the power-law correlation.}
\centering
\begin{tabular}{c c c c c c}
\hline \hline
Correlation & $\rho$ & P$(\rho)$ & $p$(fit) & $p$(CB) & $p$(FB)\\
\hline
$E'_p-E_{iso}$&  +0.87 & $\sim~0$  & $0.54\pm .01$ & $1/2$ & 1  \\
$t'_b-(E'_p\,E_{iso})$ &-0.74&$1.3\times 10^{-10}$&$-0.58\pm 
.04$&-1/2 & \\
$t'_b -E'_p$ &-0.63& $ 1.0\times 10^{-6}$ & $-1.62\pm .04$&-3/2 &-1 
\\
$t'_b-E_{iso}$ &-0.49& $4.5\times 10^{-4}$ & $-0.69\pm .06$ &-3/4 
&-1\\
\hline 
\end{tabular}
\end{table}

\begin{table}
\caption{The $\chi^2/dof$ statistic for 
the best fit  power-law correlations
between $E'_p$, $E_{iso}$ and $t'_b$ 
and for the correlations predicted by
the cannonball (CB) and collimated fireball (FB) models.}
\centering
\begin{tabular}{c c c c c c}
\hline \hline
Correlation & dof &  & best fit  & CB & FB\\
\hline
$E'_p-E_{iso}$         & 108 &  & 19.2  &20.2 & 99.7  \\
$t'_b-(E'_p\,E_{iso})$ &  53 &  & 26.9  &27.5 &  ---  \\
$t'_b -E'_p$           &  53 &  & 22.3  &23.1 & 27.6  \\
$t'_b-E_{iso}$         &  53 &  & 23.0  &24.6 & 30.1  \\
\hline
\end{tabular}
\end{table}

\appendix
\section{Ballistic conical shells}
Consider the deceleration of a highly relativistic conical shell
of a solid angle $2 \pi\, (1-cos\theta_j)\approx \pi\,\theta_j^2$  
that expands  radially  and decelerates by
sweeping in the medium in front of it. Assuming a plastic collision
and neglecting radiation losses, relativistic energy-momentum
conservation, $d(M\, \gamma)=0$, can be written as,
\begin{equation}
{M_0\,\gamma_0\over\gamma}\, d\gamma
=-\gamma^2\,\pi\,\theta_j^2\,n_b\,m_p\,R^2\,dR\,,
\label{EC}
\end{equation}
where $M(t)=M_0\, \gamma(0)/\gamma$ is the mass of the jet,
$M_0=M(0)$, $\gamma_0=\gamma(0)$, $n_b$ is the
constant baryon density of the external medium, and $R(t)$ is the radius
of  the conical shell. Eq.~(A.1) yields
\begin{equation}
[R^3-R_0^3]= {3\, M_0\,\gamma_0 \over 2\,\pi\,\theta_j^2\,n\,m_p}\,
              \left[{1\over \gamma^2}-{1\over \gamma_0^2}\right].   
\label{Rgamma}
\end{equation}
In the conical fireball model, ordinary GRBs have
$\gamma_0^2 \,\theta_j^2 \gg 1 $, whereas
$\gamma(t'_b)\theta_j\approx 1 $ at the break-time
$t'_b$ in the GRB rest frame 
corresponding to a break-time $t_b$
in the observer frame.
Consequently, Eq.~(A.1) implies that
the radius of the conical shell at $t'_b$ is given by
\begin{equation}
R_b=R(t'_b)\approx \left[{3\,E_k\over 2\,\pi\, n\, 
m_p\,c^2}\right]^{1/3}\,,
\label{Rb}
\end{equation}
where $E_k\simeq M_0\,\gamma_0\, c^2$ is the kinetic energy of the conical    
ejecta.
Eq.~(A.1) yields also the asymptotic behaviour $R\approx c\,t'\approx
R_b\,[\gamma(t')/\gamma_b]^{-2/3}$, which is reached already well before
$t'_b$.
Because of time aberration and cosmic expansion, the on-axis observer's
time interval $dt$ that corresponds to the time interval
$dt'=dR/c $
in the GRB rest frame is given by  $dt=(1+z)\, dt'/2\,\gamma^2$.
Consequently, $\gamma(t')$ has the asymptotic behaviour $\gamma_b\,
(t/t_b)^{-3/8}$ where
the observed on-axis break-time is given by
\begin{equation}
t_b=\int^{t_b} dt =(1+z)\,\int^{t'_b} {dt'\over 2\, \gamma^2}\approx
             {(1+z)\,\, R_b\,\theta_j^2\over 8\,c}\,,
\label{tb1}   
\end{equation}
and where we have assumed that most of the contribution to the integral 
comes
from times when $\gamma(t')=\gamma_b\,(t'/t'_b)^{-3/2}$ is already a  good
approximation.

If the total gamma-ray energy emitted in GRBs
is a constant fraction of the initial kinetic  energy
of the conical shell, $E_{\gamma}=\eta\, E_k$, which
in the conical fireball model is related to the isotropic equivalent
gamma-ray energy  $E_{iso}$
of the GRB by \footnote{
The solid angle of a conical shell is $\Omega_j= 2\,\pi\,
(1-cos\theta_j)$, which yields a beaming factor
$f_b= 2\, \pi\, (1-cos\theta_j)/4\,\pi= (1-cos\theta_j)/2$,
and not $f_b=(1-cos\theta_j)$ that is widely used in the GRB
literature.} $E_{iso}\approx 4\, E_\gamma/\theta_j^2$
then
\begin{equation}
t_b ={(1+z)\over 16\,c} \left[{3\,E_{iso}\over
           \pi\, \eta\, n\, m_p\, c^2} \right]^{1/3}\, \theta_j^{8/3}.
\label{tb2}  
\end{equation}
Eqs.~(A.4, A.5) were used
by Sari et al.~(1999) to represent conical jets with lateral expansion.

A distant observer sees only the beamed radiation from an area $R^2\,
\pi/\gamma^2$ along the line of sight of a spherical shell or a conical
shell. Consequently, as long as $\gamma(t)>1/\theta_j$ the observed
afterglows from a conical fireball or an isotropic fireball have the same
visible area. Beyond the break the visible area of a spherical fireball
continues to be $\approx R^2 \pi/\gamma^2$, while that of a conical shell
becomes $\approx R^2\, \pi\, \theta_j^2$. Hence the light-curve of the
afterglow of the conical fireball beyond the break is steeper by their
ratio $\gamma^2\,\theta_j^2\approx (t/t_b)^{-0.75}$, where we used the
asymptotic behaviour $\gamma(t)=\gamma_b\ (t/t_b)^{-3/8}$ in a constant
density environment and $\gamma_b=1/\theta_j$. This steepening of the
power-law decline by $\Delta \alpha=0.75$ across the break independent of
$\beta$ is different from that derived by Sari et al.~(1999) for a
spreading jet.
The smooth transition between the pre-break and post-break power laws can
also be derived more rigorously: Relativistic beaming and Doppler boosting
modulates the observed emission from every point on the conical shell by a
factor $\delta^{1+\Gamma}$ where $\delta=1/\gamma\,(1-\beta\, cos\theta)$
is the Doppler factor on the shell at an angle $\theta$ relative to the
line of sight to the observer, $\beta=v/c$  and $\Gamma$ is the photon 
spectral index of
the radiation. For isotropic medium, isotropic conical shell and isotropic
expansion, this is the only dependence of the received radiation on the
line of sight to the observer. Consequently, the observed
energy-flux (at a given energy) of photons 
emitted simultaneosly by the conical shell 
is modulated by the factor
\begin{equation}
I(\gamma,\theta_j)=2\pi\int \delta^{\Gamma+1} dcos\theta
            ={2\,\pi \over \beta\,\gamma\,\Gamma}
\left[{1\over
(1-\beta)^\Gamma}-{1\over (1-\beta\,cos\theta_j)^\Gamma}\right].
\label{I2}    
\end{equation}
The difference in arrival times of photons emitted simultaneously from
the
conical shell was ignored in the above analytical estimates of the
break-time and the spectral index change across the break. The spread in
arrival times has no effect on $E_{iso}$, it is $\Delta t\approx
R/2\,c\,\gamma^2$ before the break and $\Delta t\approx R\,\theta_j^2/c$
after the break. Thus, $\Delta t$ is roughly 4 times larger than $t$
before the break but smaller than $t$ by a factor
$4\,\gamma^2\,\theta_j^2\ll 1$ at late times. The spread in arrival time 
that has a negligible effect on $E_{iso}$ and $\Delta \alpha$ cannot be
neglected in estimating $t_b$. The same conclusion is valid also for the
effects of off-axis viewing when the viewing angle $\theta$ is not
negligible compared to $\theta_j$. Generally, the effects of off-axis  
viewing and the spread in arrival time require numerical integrations
(e.g., van Eerten and MacFadyen~2012) and make the widely used simple
relation, Eq.~(A.4) a very rough estimate.

If the typical circumburst region of LGRBs is the wind region of a Wolf 
Rayet star that blows a constant wind, than its density profile is
$\rho =\rho_0\,R_0^2/R^2=\dot{M}/4\,\pi\,R^2\, V $ where the typical mass-loss 
rate $\dot{M}\sim 10^{-4}\, M_\odot {\rm y^{-1}}$ and the typical wind 
velocity $V\sim 1000\, {\rm km\, s^{-1}}$ yield $\rho_0\, R_0^2= 
5\times 10^{11}\, {\rm g\, cm^{-1}}$. The replacement of 
$n_b\, m_p\, R^2$ with $\rho_0\, R_0^2$ in the Eq.~(A.1),
and repetition of  the derivations of the break-time  correlations for a 
constant density, yield for $\gamma\,\theta=1$   
 \begin{equation}
R_b\approx {E_k\over 2 \pi\, \rho_0\, R_0^2\, c^3}
\label{Rbw}
\end{equation}
i.e., a typical $R_b\approx 3.5\times 10^{16}$ independent of 
$\theta$, and 
\begin{equation}
t'_b\approx {\theta^2\, E_k\over 8\,\pi\, \rho_0\, R_0^2\, c^3}=
{\eta\, E_k^2\over 2\,\pi\, \rho_0\, R_0^2\, c^3}\, {1\over E_{iso}}.  
\label{tbw}
\end{equation}
Hence, the break-time power-law correlations for ISM and 
wind-like density profiles have identical 
power-law indices, i.e.,  $t'_b \propto 1/E_{iso} \propto 1/E'_p$.

\section{Ballistic cannonballs}
In the cannonball model the electrons that enter the CB are
Fermi accelerated and cool rapidly by synchrotron radiation (SR).  This SR
is isotropic in the CB's rest frame and has a smoothly broken power-law
spectrum with  a characteristic bend/break frequency,
which is the typical synchrotron frequency radiated by the interstellar
medium (ISM) electrons that enter  the CB at time $t$ with a relative
Lorentz factor $\gamma(t)$.
In the observer frame, the emitted photons are beamed
into a narrow cone along the CB's direction of motion
by its highly relativistic bulk motion, their arrival times
are aberrated and their energies are boosted by
its  bulk motion Doppler factor $\delta$ and redshifted by the cosmic
expansion during their  travel time to the observer.
For the X-ray band that is well above the break frequency, the CB
model yields the  spectral energy density  (see, e.g., Eq.~(26) in Dado
et al.~2009a),
\begin{equation}
F_{\nu} \propto  n^{(\beta_X+1)/2}\,
[\gamma(t)]^{3\,\beta_X-1}\, [\delta(t)]^{\beta_X+3}\, \nu^{-\beta_X}\,.
\label{Fnu}
\end{equation}
For a CB of a baryon number $N_{_B}$, a constant or  slowly
expanding radius $R$ and an initial Lorentz
factor $\gamma_0=\gamma(0)\gg 1$, which propagates in an ISM of {\it a 
constant density} $n_b$, relativistic energy-momentum conservation
yields the deceleration law (Dado et al. 2009b and references therein) 
\begin{equation}
\gamma(t) = {\gamma_0\over [\sqrt{(1+\theta^2\,\gamma_0^2)^2 +t/t_0}
          - \theta^2\,\gamma_0^2]^{1/2}}\,,
\end{equation}
where $t_0\!=\!{(1\!+\!z)\, N_{_{\rm B}}/ 8\,c\,n\,\pi\, R^2\,\gamma_0^3}\,.$
As long as  $t< t_b\!=\!(1\!+\!\gamma_0^2\theta^2)^2\,t_0$,
$\gamma(t)$ and $\delta(t)$
change rather slowly with $t$, which generates the plateau phase of
$F_\nu(t)$ of canonical X-ray AGs that was predicted by the CB model
(see, e.g., Dado et al.~2002, Figs.~6, 27-33) and later observed with
Swift (Nousek et al.~2006).
For $t\gg$  $t_b$,
$\gamma(t)\!\rightarrow\!\gamma_0 (t/t_b)^{-1/4}$,   
$[\gamma(t)\theta]^2$ becomes $\ll\! 1$ and  $\delta\!\approx\!
2\,\gamma(t)$,  which  result in a post-break  power-law decline
\begin{equation}
F_\nu(t)\propto \nu^{-\beta_X}\,t^{-(\beta_X+1/2)}\,.
\label{latefnu}
\end{equation}
Thus, in the CB model, the asymptotic post-break decline
of the X-ray afterglow of a single CB in an ISM environment
satisfies the closure relation 
$\alpha_X=\beta_X+1/2=\Gamma_X-1/2$   
(or $\alpha_X=\beta_X=\Gamma_X-1$ for  a shot-gun configuration 
of CBs) independent of the pre-break behaviour,
which strongly depends on viewing angle (e.g., Dado and 
Dar 2012a).

\noindent
{\bf Acknowledgement}: We thank an anonymous referee for useful 
comments and suggestions.

\end{document}